\documentclass{aa}  

\usepackage{graphicx}
\usepackage{txfonts}
\usepackage{lscape}
\usepackage{multirow}
\usepackage{longtable}
\usepackage{natbib}
\begin{document}

\title{Satellite Impact on Astronomical Observations Based on Elliptical Orbit Model}

  \subtitle{}

   \author{Tianzhu Hu
          \inst{1,2}
          \and
          Yong Zhang\inst{1,2,4}
          \and
          Xiangqun Cui\inst{1,2}
          \and
          Zihuang Cao\inst{4}
          \and
          Kang Huang\inst{1,2,3}
          \and
          Jingyi Cai\inst{1,2}
          \and
          Jun Li\inst{1,2}
          \and
          Tong Zhou\inst{1,2,3}
          }

   \institute{Nanjing Institute of Astronomical Optics \& Technology, Chinese Academy of Sciences, Nanjing 210042, China\\
              \email{yzh@niaot.ac.cn;xcui@niaot.ac.cn}
         \and
             CAS Key Laboratory of Astronomical Optics \& Technology, Nanjing Institute of Astronomical Optics \& Technology, Nanjing 210042, China
         \and
         	 University of Chinese Academy of Sciences, Beijing 100049, China
         \and
         	 National Astronomical Observatories, Chinese Academy of Sciences, Beijing 10049, China
             }

   \date{Received :; accepted :}

 
 \abstract
   {Space-based and ground-based telescopes have extensively documented the impact of satellites on astronomical observations. With the proliferation of satellite mega-constellation programs, their influence on astronomical observations has become undeniable. It is crucial to quantify the impact of satellites on telescopes. To address this need, we have enhanced the circular orbit model for satellites and introduced a methodology based on two-line element (TLE) orbit data. This involves constructing a satellite probability distribution model to evaluate the impact of satellites on telescopes. Using our method, we assessed the satellite impact on global observatories. The results indicate that the regions most severely affected by satellite interference currently are those near the equator, with latitudes around $\pm50$ and $\pm80$ degrees experiencing the most significant impact from low Earth orbit satellites. Furthermore, we validated the reliability of our method using imaging data obtained from the focal surface acquisition camera of the LAMOST telescope.}


   \maketitle
%

\section{Introduction}

Since the launch of the Iridium satellite constellation comprising 66 satellites in 1997 \citep{1998JBAA..108..187J}, and up to the current date of October 26, 2023, with Starlink having deployed over 4900 satellites, the impact of these satellite constellations on astronomical observations has gained increasing attention from astronomers and astronomy enthusiasts \citep{2023NatAs...7..252B,2023NatAs...7..237F}. Proposals for additional satellite constellations similar to the Starlink project, such as OneWeb and Amazon's Project Kuiper, are on the rise. Projections suggest that by 2030, the number of low Earth orbit satellites dedicated to communication will surpass 100,000 \citep{2022NatAs...6..428L}. 

Satellites produce bright trails on images captured by ground-based and space telescopes due to sunlight reflection, leading to contamination of imaging and spectral data \citep{2023NatAs...7..262K, 2022ApJ...924L..30M, 2023MNRAS.525L..60K}. The diffuse reflection of space debris resulting from satellite collisions contributes to an increase in the brightness of the night sky, reducing imaging contrast \citep{2021MNRAS.504L..40K, 2023NatAs...7..252B}.  Communication satellites also introduce significant radio frequency interference for radio telescopes \citep{2023A&A...676A..75D, 2023A&A...677A.141L, 2023A&A...678L...6G}. Short-term occultations of observed targets by satellites can lead to inaccurate photometry \citep{2022A&A...667A..45G}. 

The initial Starlink V1.0 satellites had a visual magnitude of 5.1, which decreased to 7.1 after the application of low-reflectance coatings. The final Starlink VisorSats have a visual magnitude of 6.0 \citep{2022MNRAS.516.1502H}. Concurrently, the median visual magnitude of OneWeb satellites is recorded at 8.23 \citep{2023PASP..135i5003K}. In their pursuit of enhanced communication signals, SpaceMobile designed the BlueWalker 3 satellite with an expansive size of 64 square meters, resulting in a visual magnitude of 0.2 \citep{nandakumar2023high}. 

In the face of the escalating severity of satellite interference, effective measures have been proposed to reduce its impact. \citet{2022MNRAS.509.1848O} achieved satellite transit forecasting using ELT data to assist in observational planning and avoidance of satellites. \citet{2022ApJ...941L..15H} introduced a method that does not rely on precise satellite orbit prediction. It involves constructing a satellite distribution map and devising observational strategies to avoid regions with a high satellite density.

Meanwhile, various models have been proposed for quantifying the impact of satellites on astronomical observations. \citet{2020ApJ...892L..36M} introduced distinct categories of satellites in orbit and assumed an evenly spaced satellite distribution along the orbital paths. They utilized this assumption to develop a transient satellite distribution model and analyzed the number of illuminated Starlink satellites at varying latitudes, times of the year, and times of the night. In a related work, \citet{2020A&A...636A.121H} introduced the concept of a uniform satellite distribution model, wherein satellites are uniformly dispersed across the Earth's surface. They investigated how the count of observable satellites is affected by solar latitude and estimated the brightness of satellites. Building upon this work, \citet{2022A&A...657A..75B} refined the satellite uniform distribution model, considering circular orbits with a certain inclination angle for the satellites, enabling a more precise estimation of both the probability distribution of satellites and their brightness.

To analyze the impact of all satellites on observations, it is necessary to consider not only low Earth orbit satellites but also high Earth orbit and highly elliptical orbit satellites. A satellite distribution probability model based on elliptical orbits is required for this purpose. In our research, we optimized the circular orbit model by considering satellite elliptical orbits. We constructed a satellite probability distribution model based on these elliptical orbits. Utilizing real satellite distribution data, we provided the probability of observing satellites from various ground stations at different viewing angles. Additionally, we conducted a reliability check of our approach by comparing it to observation results. We used imaging data obtained from LAMOST's acquisition cameras \citep{2012RAA....12.1197C,2023MNRAS.525.3541H} to analyze the probability of satellite observations from the Xinglong Observatory, validating the effectiveness of our method.

The structure of this paper is organized as follows: The first section provides an overview of the data sources we utilized in our research, including satellite orbit data and data from the LAMOST guide cameras. The second section delves into our satellite probability distribution model, which is based on elliptical orbits. We explain the methodology and equations used to construct this model. The third section focuses on analyzing the impact of satellites on observations at the Xinglong Observatory, specifically using data from the LAMOST acquisition cameras. We compare the observed results with the predictions from our satellite probability distribution model. Finally, we summarize some conclusions based on the discussion part.

\section{Data}

In this section, we provide the data pertinent to the development and validation of the probability model for satellite distribution. Data encompasses information about satellite orbital and the LAMOST acquisition camera.
 
\subsection{Satellite orbital data}
 
\begin{figure}
   \centering
   \includegraphics[width=\hsize]{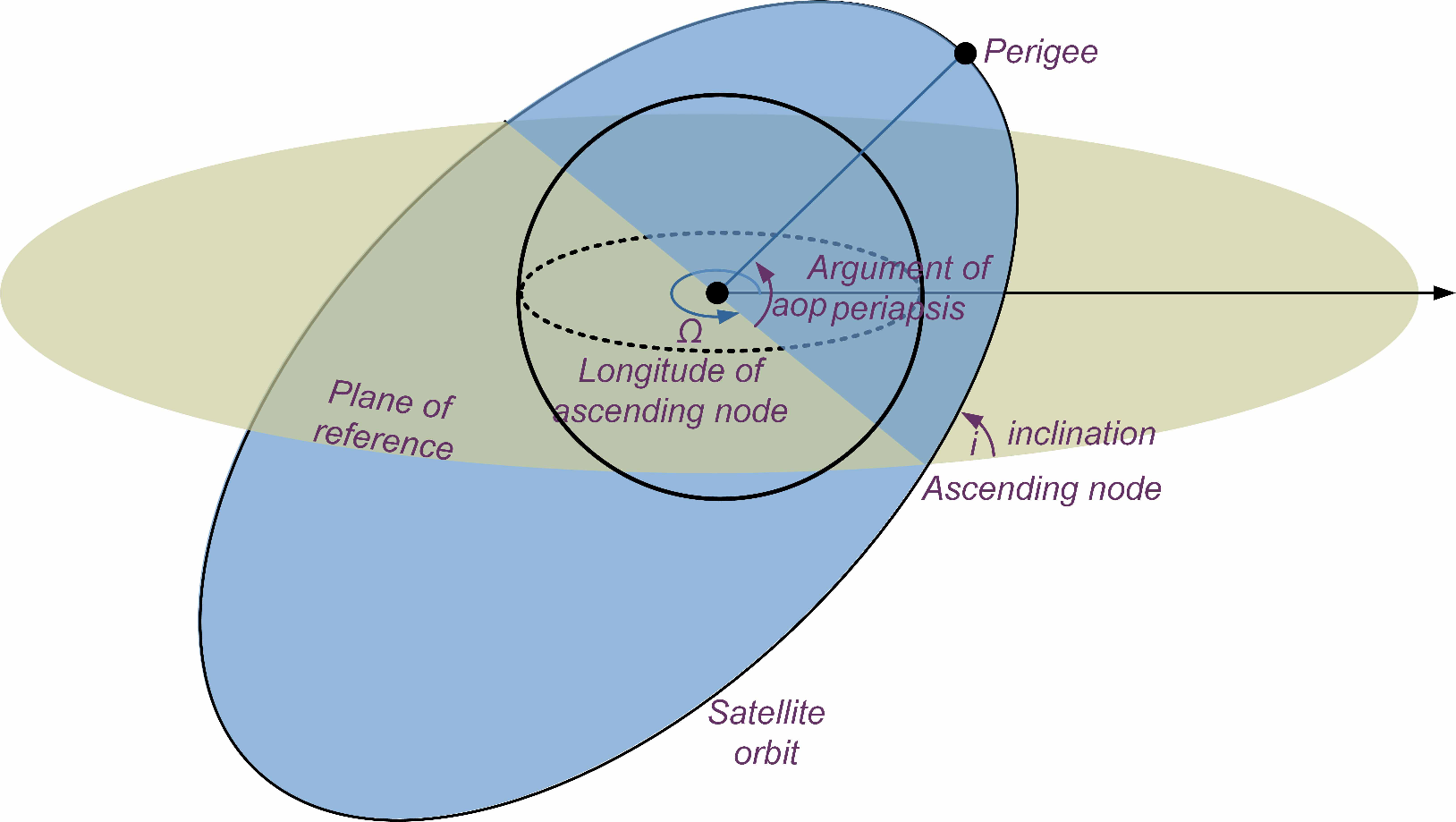}
      \caption{To determine the orbit of a satellite, several parameters are required. These include the orbital inclination, which is the angle between the satellite's orbital plane and the Earth's equatorial plane. Additionally, the satellite's right ascension of the ascending node (RAAN) denotes the point where the satellite crosses from the southern hemisphere to the northern hemisphere while rotating counterclockwise with respect to the right ascension. The argument of perigee signifies the angle between the line connecting the Earth's center to the nearest point on the orbit (perigee) and the line connecting the Earth's center to the ascending node. Other essential parameters include the eccentricity and semi-major axis of the orbit.}
         \label{aa_fig1}
 \end{figure}

A satellite's orbit can be determined by six parameters, commonly including orbital inclination, right ascension of the ascending node, orbital eccentricity, argument of perigee, perigee altitude, and true anomaly, as illustrated in Fig.~\ref{aa_fig1}. Satellite orbit data is stored as TLE files with three lines, each containing 69 characters. From these files, we extract the necessary data, which includes the orbital inclination ($i$), argument of perigee ($ aop $), and orbital eccentricity ($e$). Additionally, the number of orbits a satellite completes in a day ($n$) is provided, enabling the calculation of the satellite's orbital period ($T$), as well as the semi-major axis ($a$) and semi-minor axis ($b$) of the orbit. Where the 
 
 \begin{equation}
 	T=\frac{24\times60\times60}{n}
 \end{equation}
 
 \begin{equation}
 	a={(\frac{GMT^2}{4\pi^2})}^{1/3}
 \end{equation}
 
 \begin{equation}
 	b=a\times\sqrt{1-e^2}
 \end{equation}
 
We downloaded orbit data for all in-orbit satellites from the \href{space-track website}{https://www.space-track.org}, amounting to a total of 25,606 objects. This dataset includes geostationary satellites, medium Earth orbit satellites, low Earth orbit satellites, and high elliptical orbit satellites, among others. After removing satellite debris, there remained 14,214 objects, with 9,381 of them identified as low Earth orbit satellites. The Tab.~\ref{table_1} provides the number of satellites for prominent satellite mega-constellations.

\begin{table}
\caption{The satellite constellations with more than 100 satellites and their respective satellite counts are as follows}             
\label{table_1}      
\centering                          
\begin{tabular}{c c}        
\hline\hline                 
 The satellite constellations & The number of satellite \\    
\hline                        
   STARLINK & 4621 \\      
   ONEWEB & 634 \\
   FLOCK & 185 \\
   GLONASS & 140 \\
   YAOGAN & 120 \\
   IRIDIUM & 109 \\
   LEMUR & 104 \\
\hline                                   
\end{tabular}
\end{table}
 
\subsection{LAMOST aquisition camera data}

\begin{figure}
   \centering
   \includegraphics[width=\hsize]{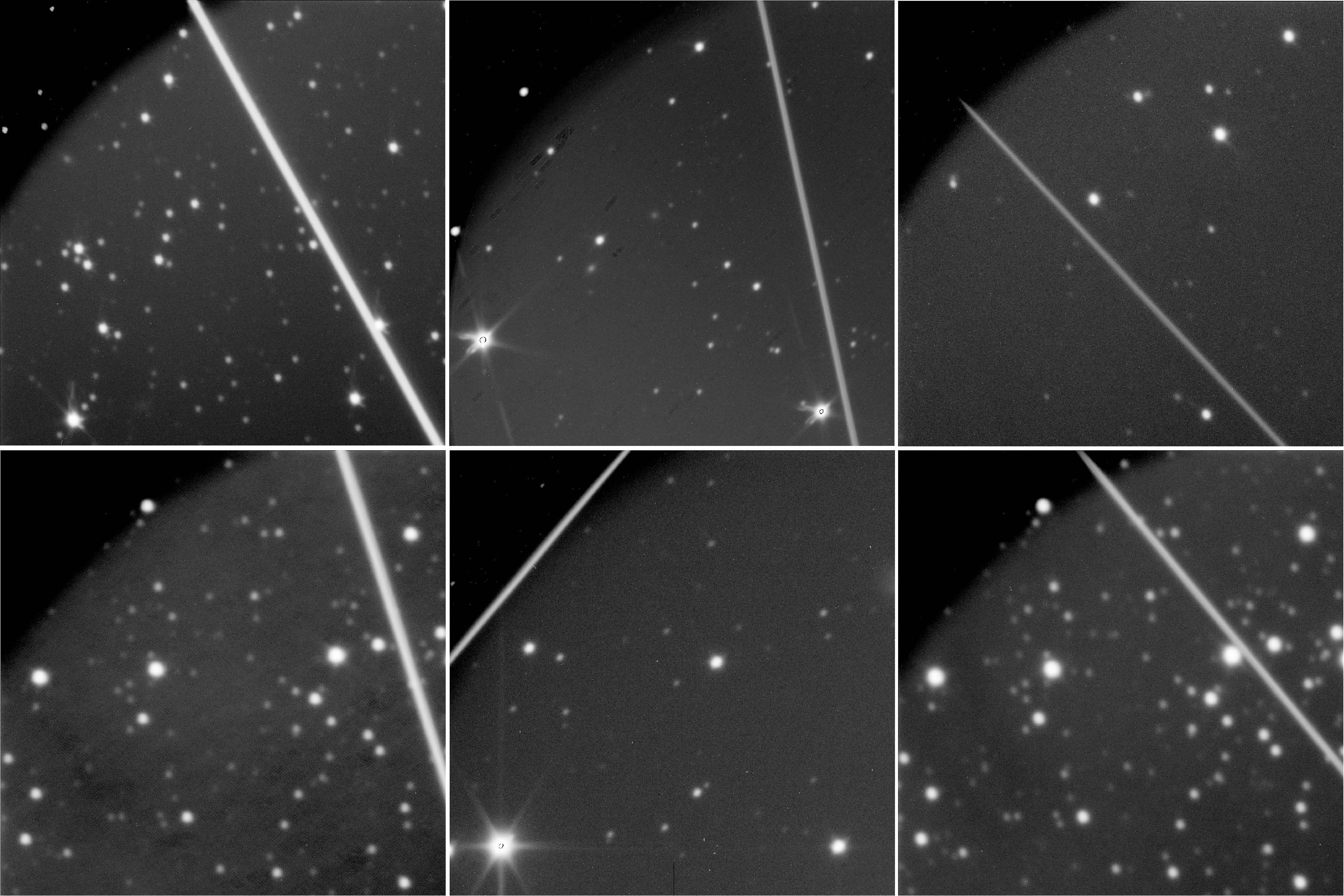}
      \caption{The images depict satellite tracks recorded on the focal surface of the acquisition cameras of the LAMOST. The images in the bottom-left and bottom-right corners document the trajectories of two satellites obtained during observations in the same celestial region.}
         \label{aa_fig2}
 \end{figure}

LAMOST is a fiber-fed spectroscopic telescope, consisting of optical paths Ma and Mb, along with a focal surface equipped with 4,000 optical fibers. It boasts a 5-degree field of view for wide-field sky surveys, making it particularly susceptible to the impact of satellites. The focal surface accommodates eight acquisition cameras, positioned at 1.5-degree and 2.5-degree field positions, recording the trajectories of satellites. Utilizing satellite trail in the imaging data, we can estimate their influence on the sky survey telescope. Each camera comprises $1024\times1024$ pixels, with each pixel measuring 24 micrometers, corresponding to a field of view of 0.025 arcseconds. Fig.~\ref{aa_fig2} presents a sample image influenced by satellites as captured by the acquisition cameras.

\section{Method}

In this section, we describe the extension of the method proposed by \citet{2022A&A...657A..75B} to elliptical orbits, enabling ELT data, including data for high eccentricity satellite orbits.

We constructed a geographic coordinate system with the Earth's center as the origin, and the satellite's coordinates can be expressed as $(lat, lon, R)$. In this representation, ($lat$, $lon$) denotes the latitude and longitude of the intersection point between the satellite-to-Earth center line and the Earth's surface, while R represents the distance from the satellite to the Earth's center. The orbital trajectory of a satellite can be expressed by the elliptical equation:

\begin{equation}
	R=\frac{a(1-e^2)}{1+ecos(\theta-aop)}
\end{equation}

\begin{figure}
   \centering
   \includegraphics[width=\hsize]{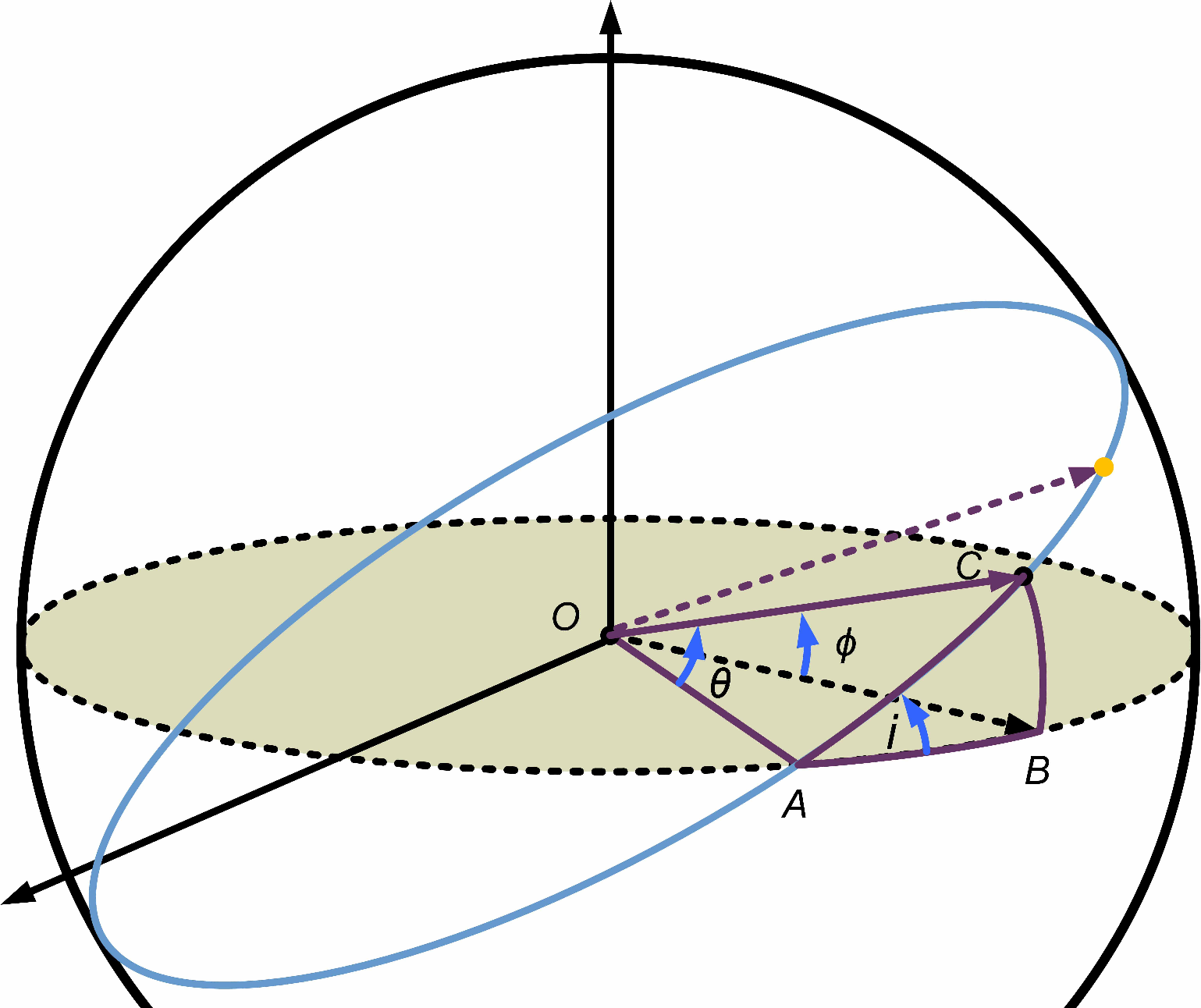}
      \caption{In the geographic coordinate system, the projection of the ascending node on the celestial sphere is denoted by point A, and the projection of the satellite's position is point C. In this system, $i$ denotes the orbital inclination, and $\phi$ represents the satellite's latitude.}
         \label{aa_fig3}
 \end{figure}

Here, $\theta$ represents the angle between the line connecting the satellite to the center of the Earth and the line connecting the ascending node to the Earth's center. The variable $aop$ stands for the argument of perigee. We can express $\theta$ using the satellite's latitude $\phi$ in the celestial coordinate system, as illustrated in  Fig.~\ref{aa_fig3}. In the spherical triangle $O-ABC$, the angle $\theta$ of the ellipse is related to the latitude in the geographic coordinate system, as shown below.

\begin{equation}
	\frac{sin\theta}{sin90^{\circ}}=\frac{sin\phi}{sin{i}}
	\label{equ5}
\end{equation}

Hence, the orbit's altitude can be represented by the satellite's latitude. The probability of a satellite being within a range of $d\theta$ angles can be calculated as the ratio of the area subtended by $d\theta$ degrees between the elliptical orbit and the total area of the ellipse, assuming the satellite fully complies with Kepler's Second Law.

\begin{equation}
	dp=\frac{R^2d\theta}{\pi ab}
	\label{equ6}
\end{equation}

Taking the differential of Equation~\ref{equ5}, we obtain: $d\theta=\frac{cos\phi}{cos\theta sini}d\phi$. Substituting this into Equation~\ref{equ6} results in

\begin{equation}
	\frac{dp}{d\phi}=\frac{R^2}{\pi ab}\frac{cos\phi}{cos\theta sini}
\end{equation}

Taking into account that the same $\phi$ corresponds to two different $\theta$, denoted as $\theta_1$ and $\theta_2$, with corresponding satellite altitudes 
$R_1$ and $R_2$, the above equation should be modified to:

\begin{equation}
	\frac{dp}{d\phi}=\frac{{R_1}^2}{2\pi ab}\frac{cos\phi}{cos\theta_1sini}+\frac{{R_2}^2}{2\pi ab}\frac{cos\phi}{cos\theta_2sini}
\end{equation}


Considering that, in general, within a given orbital plane, there will be multiple satellites with the same orbital inclination, and accounting for the Earth's rotation, the right ascension of the ascending node for satellites will continuously change. Consequently, the relative positions of satellites will also vary. As a simplifying assumption, we consider that the probability of a satellite at a given latitude is equal for different longitudes. Thus, we can obtain the probability of a satellite in a unit solid angle as:

\begin{equation}
\begin{split}
	p_{sat} & =\frac{dp}{d\mathrm{\Omega}}=\frac{dp}{d\phi\times2\pi\times cos\phi}\\
	& =\frac{{R_1}^2}{4\pi^2ab}\frac{1}{cos\theta_1sini}+\frac{{R_2}^2}{4\pi^2ab}\frac{1}{cos\theta_2sini}
\end{split}
\label{equ9}
\end{equation}

According to the formulas presented above, we can calculate the probability distribution of satellites corresponding to various longitudes and latitudes.

\section{Result and Discussion}

In this section, we present the key conclusions of our method and validate the model using observational data.

\subsection{Satellite Probability Distribution}

\begin{figure*}
   \sidecaption
   \includegraphics[width=12cm]{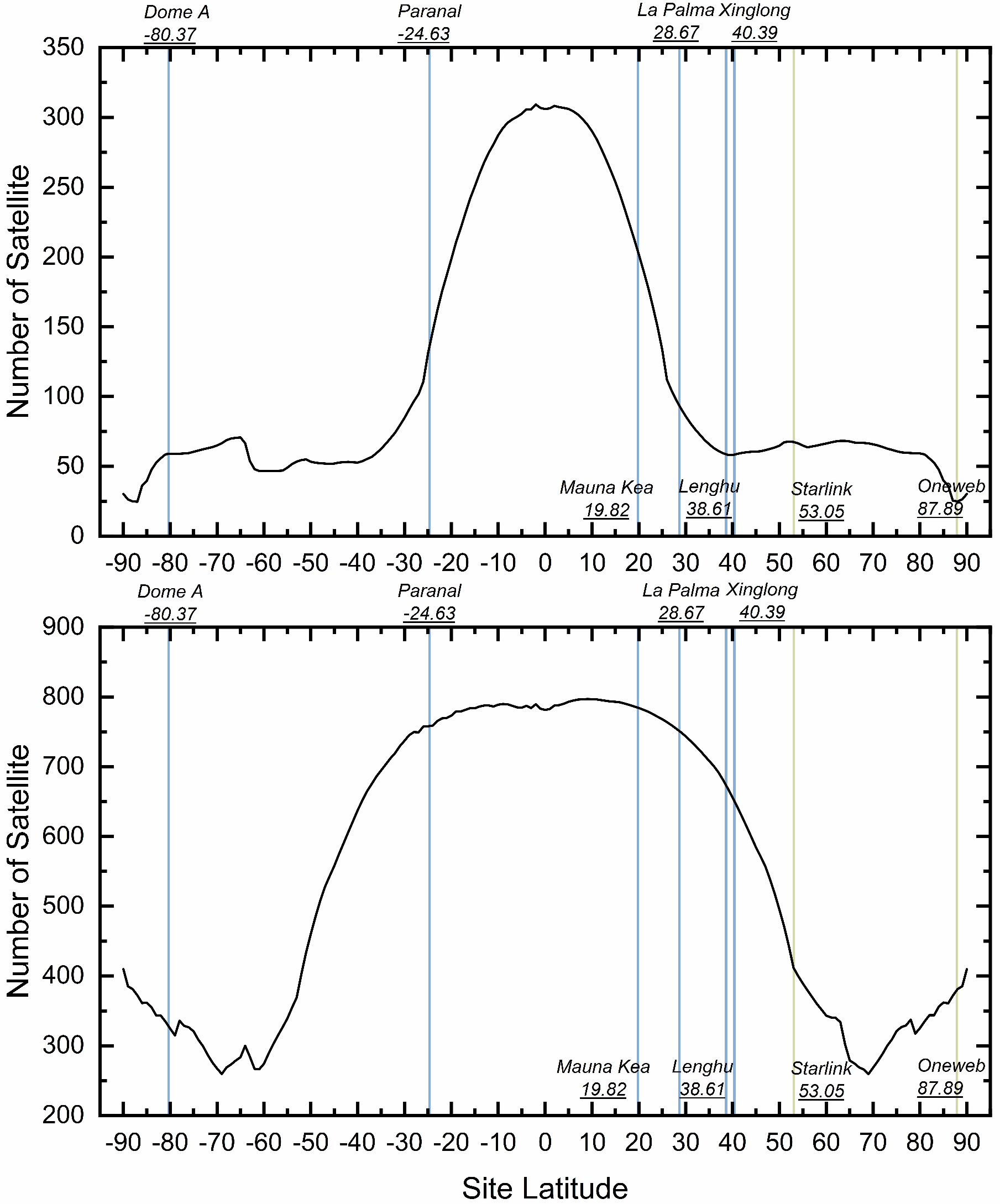}
      \caption{The image illustrates the observable satellite count at different latitudes, considering all satellites, with the equatorial region being most severely affected. The upper figure depicts the count of observable satellites within a 30-degree field of view centered on the zenith at the observatory, while the lower figure represents the count of observable satellites within a 60-degree field of view centered on the zenith at the observatory.}
         \label{aa_fig4}
\end{figure*}

Equation~\ref{equ9} addresses the probability distribution for an individual satellite, taking into account a total of N satellites in orbit. The overall observation probability signifies the count of observable satellites, necessitating the summation of the observation probabilities for each of the N satellites. The number of satellites that can be observed for an telescope with field of A with an exposure time of $t_{exp}$ can be expressed as:

\begin{equation}
	N_{sat}= \sum_N \sum_{\substack {i = 1,2}} \rho_{sat}(R_i)\left(A+L\omega_{sat}(R_i)t_{exp}\right)
	\label{equ10}
\end{equation}

The symbol $\rho_{sat}$ denotes the angular density of satellites at the telescope's location, which can be calculated through $p_{sat}$.

where

\begin{equation}
\begin{split}
		\rho_{sat} & = \sum_{\substack {i = 1,2}} p_{sat} \times \frac{d(R_i)^2\times\cos\alpha_2(R_i)}{R_i^2\times \cos\alpha_1(R_i)}\\
		& = \frac{{R_1}^2}{4\pi^2ab}\times\frac{1}{cos\theta_1sini}\times\frac{{d(R_1)}^2cos\alpha_2(R_1)}{{R_1}^2cos\alpha_1(R_1)}\\
		& +\frac{{R_2}^2}{4\pi^2ab}\times\frac{1}{cos\theta_2sini}\times\frac{{d(R_2)}^2cos\alpha_2(R_2)}{{R_2}^2cos\alpha_1(R_2)}
\end{split}
\label{equ11}
\end{equation}

\begin{equation}
	\omega_{sat} = \frac{2\pi ab}{TR_i^2}
\end{equation}

Here, $\alpha_2(R_i)$ represents the angle between the satellite-to-Earth center vector and the normal vector to the satellite orbital surface. The term $\alpha_1(R_i)$ represents the angle between the satellite-to-observation point vector and the normal vector to the satellite orbital surface, $d$ is the distance between the satellite and the observation point.

We present a detailed calculation of $\cos\alpha_1$, $\cos\alpha_2$, and $d$ in Appendix \ref{sec_app}, and $\alpha_1$, $\alpha_2$ and $d$ are shown in Fig~\ref{aa_fig8}.

\subsection{Satellite Possibility Map}

\begin{figure*}
   \sidecaption
   \includegraphics[width=12cm]{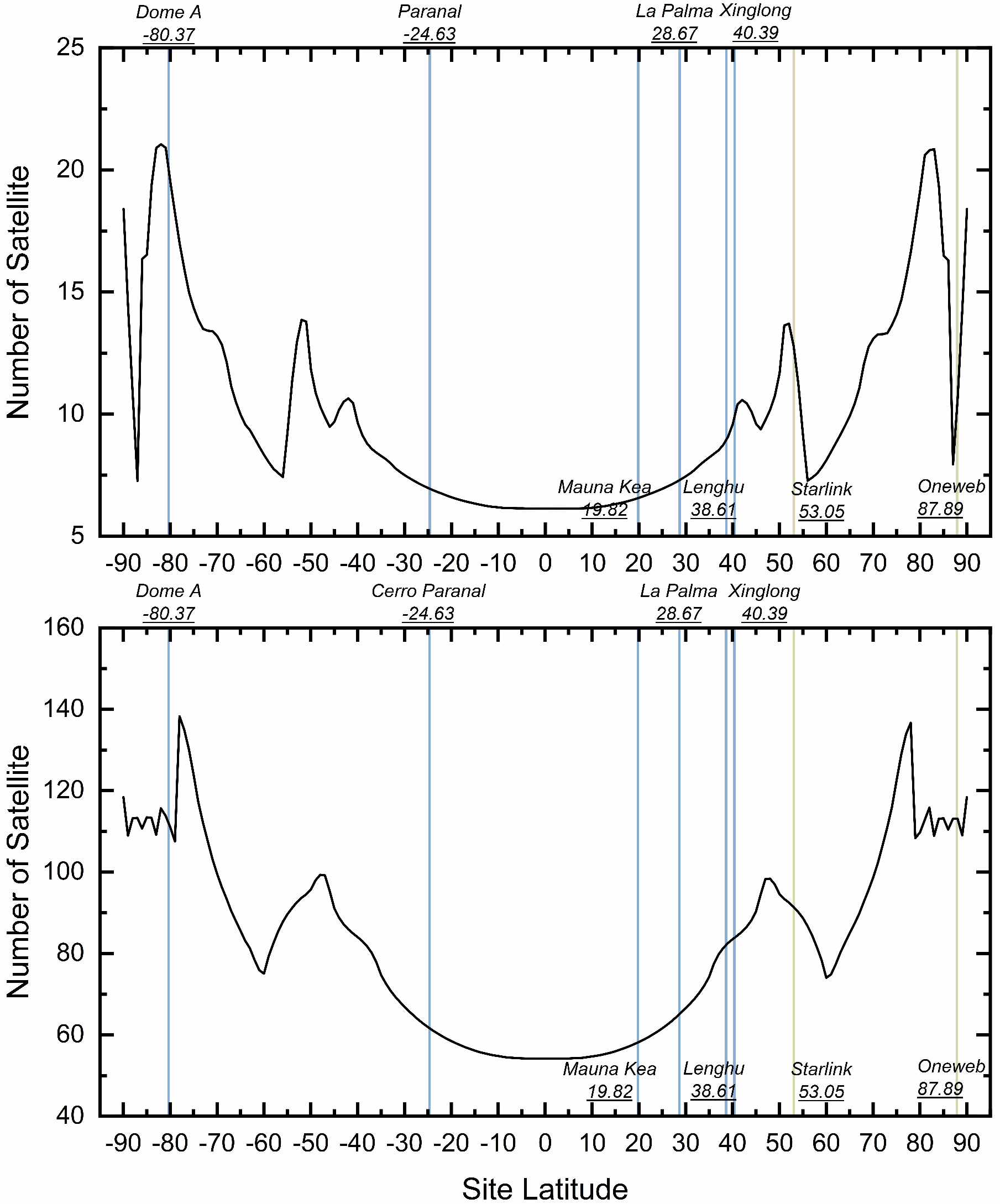}
      \caption{The image reflects the results considering only low-earth orbit satellites, with the impact being most severe around latitudes of $\pm50$ degrees and $\pm80$ degrees, consistent with the orbital inclinations of Starlink and OneWeb satellites. The upper figure depicts the count of observable satellites within a 30-degree field of view centered on the zenith at the observatory, while the lower figure represents the count of observable satellites within a 60-degree field of view centered on the zenith at the observatory.
              }
         \label{aa_fig5}
\end{figure*}

The $\rho_{sat}$ can serve as an indicator of the impact of satellites on the observation site. Considering the intricate nature of a satellite's orbital shell, typically consisting of two rotationally symmetric surfaces, we calculate the heights $R_1$ and $R_2$ of these two orbital planes at the zenith position of the observation site. In two observation modes, where the observed elevation angle is greater than 30 degrees or greater than 60 degrees, the orbital surafce can be approximated as two spherical surfaces with radii $R_1$ and $R_2$. Under this approximation, $cos\alpha_2 = 1$.

and 

\begin{equation}
	\rho_{sat}= \sum_{\substack {i = 1,2}} p_{sat}\times\frac{d(R_i)^2}{R_i^2\times cos\alpha_1(R_i)}
\end{equation}

We integrate $\rho_{sat}$ within 30 degrees and 60 degrees near the zenith of the observation site. We performed integration using TLE data for 14,214 satellites, including all satellites. The results are summed up, as shown in Fig~\ref{aa_fig4}.

The results indicate that the maximum number of observable satellites is near the equator, potentially influenced by the presence of numerous geostationary satellites. Moreover, the number of observable satellites at elevation angles surpassing 30 degrees surpasses those at angles exceeding 60 degrees.
 
Additionally, we also computed the results considering only low-orbit satellites. We integrated the data for 9,314 low-orbit satellites, reflecting the future impact of satellites on telescope observation. The results are shown in Fig~\ref{aa_fig5}. The results indicate that future satellite impact will be particularly severe in regions around $\pm50$ degrees and $\pm80$ degrees. These correspond to the typical orbital inclinations of Starlink and OneWeb satellites. Due to the higher orbit of OneWeb satellites, more observable satellites are present in $\pm80$ regions. Comparing the integration results for low-orbit satellites with those including all satellites, it is evident that the influence of high-orbit satellites still dominates the overall impact.
 
\subsection{Observation Result}

\begin{figure*}[h]
   \centering
   \includegraphics[width=0.9\textwidth]{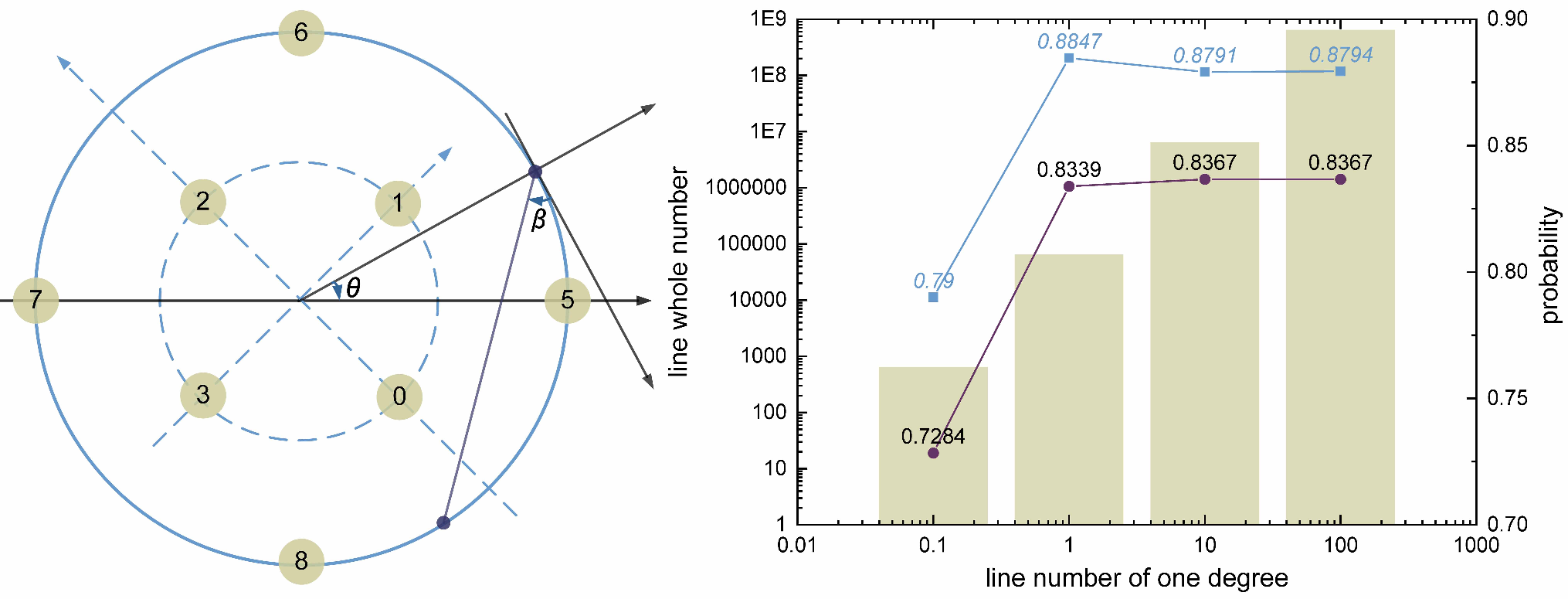}
      \caption{The left figure illustrates the configuration of LAMOST's focal surface acquisition cameras. In the right figure, the probability that the lines corresponding to different numbers do not intersect with any of the eight acquisition cameras is depicted.}
         \label{aa_fig6}
\end{figure*}

We conducted our research using data obtained from the acquisition camera of the LAMOST telescope from January 2023 to June 2023. Initially, we excluded images that couldn't be used to determine the presence of satellite trails, such as those taken during the transition between sky regions and images with camera malfunctions. The remaining dataset consists of 226,630 images, with a total imaging time of 7,295,597 seconds, of which 423 images contain individual satellite tracks.


%

The number of satellite trails recorded by the acquisition cameras is only a subset of all the satellites passing through the focal surface. By utilizing the count of satellite tracks captured by the acquisition cameras, we can estimate the total number of satellite trails acquired across the entire focal surface. This issue can be reformulated as the probability that any straight line traversing the focal surface does not intersect with the eight fixed-position small circles on the focal surface. This probability can be expressed as:

\begin{equation}
	P=\frac{\int_{0}^{2\pi}\int_{0}^{\pi}p\left(\theta,\beta\right)d\theta d\beta}{2\pi^2}
\end{equation}

Where $\theta$ and $\beta$ are as shown in Fig~\ref{aa_fig6}, $p(\theta, \beta) = 1$ when the line does not intersect the small circle, and $p(\theta, \beta) = 0$ when the line crosses the small circle. We conducted a differential simulation of $\theta$ and $\beta$, and we calculated the distance between the straight line and the midpoint of each camera, then compared the distance with the size of the camera to determine whether the line intersects with the camera. The actual imaging units of the camera are square, and we created inscribed and circumscribed circles for the squares, taking the radius of the circle as the size of the camera. The results indicate a probability range of 0.8333 to 0.8797 for the likelihood of not being imaged by any of the 8 cameras and passing through the focal surafce.

According to Equation~\ref{equ10}, the number of observed satellites during $	n$ observations can be expressed as:

\begin{equation}
\begin{split}
	N_{sat} & = \sum_{n} \sum_N \sum_{\substack {i = 1,2}} \rho_{sat}(R_i)\left(A+L\omega_{sat}(R_i)t_{exp}\right)\\
	& = n\ast(\bar{\sum_N \sum_{\substack {i = 1,2}} \rho_{sat}(R_i)}\ast A+\bar{\sum_N \sum_{\substack {i = 1,2 }}\ast\omega_{sat}(R_i)}\ast\ L\ t_{exp})\\
	& = n\ast(\bar{\rho}\ast A+\bar{\rho\omega}\ast\ L\ t_{exp})
\end{split}
\label{equ15}
\end{equation}

The averages, $\bar{\rho}$ and $\bar{\rho\omega}$, represent the mean values over $n$ times observations and can be expressed as the average of the integrated results over the observed sky regions. Considering the latitude of the LAMOST telescope is 40.3959 degrees north, the average elevation angle during observations from January to June 2023 is 51.11 degrees, with a median of 50.01 degrees, we can calculate $\bar{\rho}$ and $\bar{\rho\omega}$ of four observation scale and satellite type, the result shown in Tab.~\ref{table_2}. 

\begin{table}
\caption{Considering scenarios where the elevation angle exceeds 30 degrees and 60 degrees, we calculate the results of $\bar{\rho}$ and $\bar{\rho\omega}$ for all satellites and low Earth orbit satellites. Here, $allsat_{30}$ represents the case of considering all satellites with an observation range exceeding an elevation angle of 30 degrees, while $lowsat_{30}$ represents the case of considering only low Earth orbit satellites with an observation range exceeding an elevation angle of 30 degrees.}             
\label{table_2}      
\centering                          
\begin{tabular}{c c c}        
\hline\hline                 
 Observation Type & $\bar{\rho}$ & $\bar{\rho\omega}$ \\    
\hline                        
$allsat_{30}$   & 1.2384 & 0.0043 \\      
$allsat_{60}$  & 0.4145 & 0.0019 \\
$loworbitsat_{30}$  & 0.1610 & 0.0024 \\
$loworbitsat_{60}$  & 0.0718 & 0.0011 \\
\hline                                   
\end{tabular}
\end{table}

We also considered whether satellites could are illuminated during the observation process. It is necessary to consider whether the line connecting the satellite and the sun is blocked by the Earth, and whether the angle formed by the sun-satellite-Earth is greater than 90 degrees. If the distance from the Earth's center to the satellite along the line to the sun is greater than the Earth's radius, or if the angle formed by the sun-satellite-Earth is greater than 90 degrees, the satellite is illuminated by the sun. We considered scenarios with satellite altitudes of 1000 kilometers and 10000 kilometers in different observation mode. Through simulation, we calculated the proportion of satellites illuminated at different times within various observation ranges at the Xinglong Observatory (altitude 900 meters, longitude 117.58 degrees, latitude 40.40 degrees), where the LAMOST telescope is located.

\begin{figure*}[h]
   \centering
   \includegraphics[width=1\textwidth]{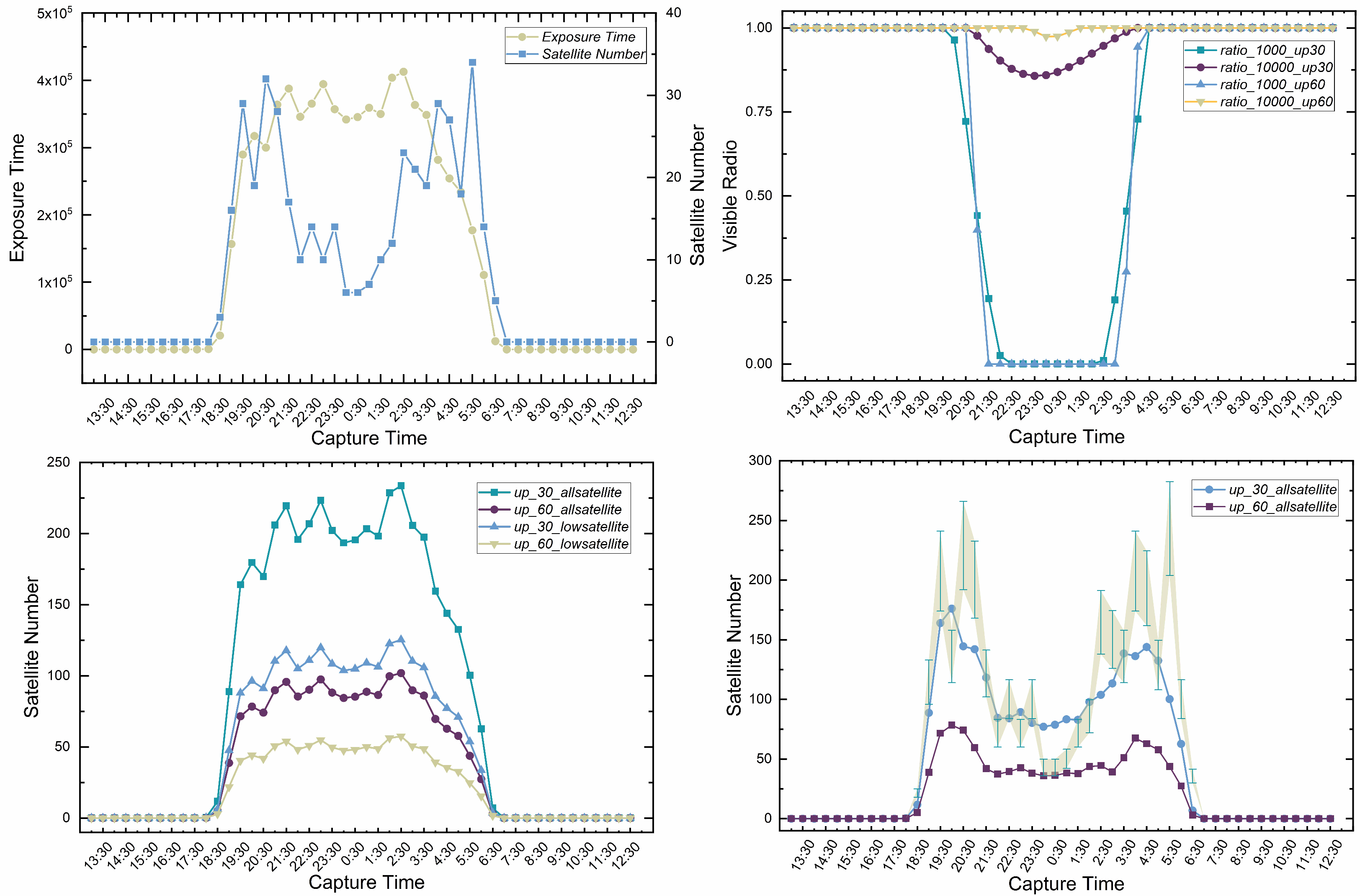}
      \caption{The top-left plot illustrates the distribution of exposure times and the corresponding times for acquiring satellite images in the observational data we utilized. The bottom-left plot presents the time distribution of observable satellites for the LAMOST telescope under the observational model rules. This includes observations for all satellites with elevation angles above 30 degrees and 60 degrees, as well as observations for near-Earth satellites with elevation angles above 30 degrees and 60 degrees. The top-right plot depicts the time distribution of satellite visibility on March 15th for satellites at altitudes of 1000 km and 10000 km, considering elevation angles above 30 degrees and 60 degrees. The bottom-right plot compares the predictions of the model, which takes into account satellite visibility, with the observed results.}
         \label{aa_fig7}
\end{figure*}

Using the integrated results of the satellite observation probability in Equation~\ref{equ15}, we obtain the number of satellites appearing during the observation time and within the observation range. We then consider whether the Sun can illuminate the satellites to determine the observable number of satellites. High-orbit satellites use illuminated probability data calculated for 10,000 kilometers, while low Earth orbit satellites use illuminated probability data calculated for 1,000 kilometers. The results are compared with the actual observational statistics, as shown in Fig~\ref{aa_fig7}. 

Taking into account that the mean zenith angle for LAMOST observations exceeds 50 degrees, the observed results should fall between observations with zenith angles greater than 30 degrees and those with zenith angles greater than 60 degrees. The results indicate good consistency between the model and observational outcomes near midnight. However, the observations during dusk and dawn show a higher discrepancy. This discrepancy may be attributed to the possibility that fragments of low Earth orbit satellites are sufficiently bright during these times, forming satellite tracks.

\section{Conclusions}

We propose an evaluation method for telescopes based on TLE data, considering the impact of not only low Earth orbit satellites but also high Earth orbit and high eccentricity satellites. The assessment, conducted using satellite data from September, reveals that, when considering all satellites, the regions near the equator are most affected due to the influence of geostationary satellites. In scenarios considering only low Earth orbit satellites, the impact is most severe around latitudes of $\pm50$ degrees and $\pm80$ degrees. This is primarily attributed to the influence of Starlink and OneWeb satellites. 

To validate the effectiveness of our method, we utilized observational data from the acquisition cameras of the LAMOST telescope for the period from January to June 2023. We conducted a statistical analysis of satellite trails captured by the guiding cameras during this period. Comparing the observational results with those considering all satellites, the findings reveal a better fit around midnight. However, there is a slight overestimation in the actual observations during the early morning and late evening. This discrepancy may be attributed to the possibility that fragments of low Earth orbit satellites are sufficiently bright during these times, forming satellite tracks. This comparison supports the validity of our method in predicting satellite interference in telescope observations.

Our method does not account for the impact of the telescope's observation limit, assuming that all satellites at different altitudes are visible. The effective aperture of LAMOST used for validation is 4.4 meters, and for telescopes with smaller apertures, the impact of high-orbit satellites would be less pronounced. Additionally, our method allows for integration over smaller regions, enabling the analysis of satellite distribution probabilities in specific observational areas, which can be useful for satellite observation schedule.

\begin{acknowledgements}

Thanks are given to the reviewer for the constructive comments and helpful suggestions. This work is supported by the National Nature Science Foundation of China  (Grant No U1931207, 12203079, 12103072 and 12073047), the Natural Science Foundation of Jiangsu Province (Grants No BK20221156 and BK20210988), and the Jiangsu Funding Program for Excellent Postdoctoral Talent. The Guo Shou Jing Telescope (or LAMOST) is a National Major Scientific Project built by the Chinese Academy of Sciences. Funding for the project was provided by the National Development and Reform Commission. LAMOST is operated and managed by National Astronomical Observatories, Chinese Academy of Sciences. 

\end{acknowledgements}

%
\bibliographystyle{aa_style} 
\bibliography{aa_2023} 
%

\begin{appendix} 
%
\section{Impact angle of observation}\label{sec_app}

In the geographic coordinate system, the satellite orbital surface is formed by the rotation of an elliptical satellite orbit around the axis of the Earth's rotation. The orbital surface equation can be expressed in terms of longitude $\theta$ and latitude $\phi$ as follows:

\begin{equation}
	r(\theta,\phi) = \frac{a*(1-e^2)}{1+e*cos(arcsin(sin\phi/sini)-aop}
	\label{a1}
\end{equation}

Equation 9 provides the angular density distribution of satellites based on the focal point (Earth's center) of the satellite orbit. To calculate the observable number of satellites during telescope observations, it needs to be transformed into a satellite probability density distribution centered around the observation point. 

According to Formula \ref{a1}, the normal vector to the orbital plane can be obtained

\begin{equation}
	\nabla r = \boldsymbol{e_r} + \frac{1}{r}\frac{\partial r}{\partial \phi}\boldsymbol{e_{\phi}}
\end{equation}

where
\begin{equation}
	\frac{\partial r}{\partial \phi} = \frac{-a(1-e^2)*e*\frac{-sin\phi}{sini}*\frac{1}{\sqrt{1-(sin\phi/sini)^2}}*\frac{cos\phi}{sini}}{1+ecos(acrsin(sin\phi)/sini-aop)^2}
\end{equation}

So, the normal vector to the orbital plane represented in Euclidean space is

\begin{equation}
	\boldsymbol{shell} = (cos\frac{1}{r}\frac{\partial r}{\partial \phi},0,Rsin\frac{1}{r}\frac{\partial r}{\partial \phi})
	\label{a4}
\end{equation}

The coordinates of the observation point and the satellite are given by ($0,\phi_{obs},R_{obs}$) and ($\theta_{sat},\phi_{sat},R_{sat}$).

The vector from the observation point to the satellite is represented as

\begin{equation}
\begin{split}
	\boldsymbol{obs} & = (Rcos\phi_{sat}cos\theta_{sat}-rcos\phi_{obs}, \\
	& Rcos\phi_{sat}sin\theta_{sat}-rcos\phi_{obs},Rsin\phi_{sat}-rsin\phi_{sat})
\end{split}
\label{a5}
\end{equation}

The vector from the Earth's center to the satellite

\begin{equation}
	\boldsymbol{earth} = (Rcos\phi_{sat}cos\theta_{sat},Rcos\phi_{sat}sin\theta_{sat},Rsin\phi_{sat})
	\label{a6}
\end{equation}

According to Formula~\ref{a4} and \ref{a5}, we can obtain 
\begin{equation}
	\cos\alpha_1 = \frac{\boldsymbol{obs}\bullet\boldsymbol{shell}}{\Vert\boldsymbol{obs}\Vert * \Vert\boldsymbol{shell}\Vert}
\end{equation}

and

\begin{equation}
\begin{split}
	d^2 & = \left(R\times sin\phi_{sat} - r\times sin\phi_0\right)^2 \\
	& + (R\times cos\phi_{sat} \times sin\theta_{sat} - r\times cos\phi_{0}\times sin\theta_{0})^2\\
	& + \left(R\times cos\phi_{sat}\times cos\theta_{sat} -r\times cos\phi_0 \times cos\theta_{0} \right)^2
\end{split}
\end{equation}

According to Formula~\ref{a4} and \ref{a6}, we can obtain 
\begin{equation}
	\cos\alpha_2 = \frac{\boldsymbol{earth}\bullet\boldsymbol{shell}}{\Vert\boldsymbol{erath}\Vert * \Vert\boldsymbol{shell}\Vert}
\end{equation}

and

\begin{equation}
\begin{split}
	R_{sat}^2 & = \left(R\times sin\phi_{sat} - r\times sin\phi_0\right)^2 \\
	& + (R\times cos\phi_{sat} \times sin\theta_{sat} - r\times cos\phi_{0}\times sin\theta_{0})^2\\
	& + \left(R\times cos\phi_{sat}\times cos\theta_{sat} -r\times cos\phi_0 \times cos\theta_{0} \right)^2
\end{split}
\end{equation}

 \begin{figure}
   \centering
   \includegraphics[width=\hsize]{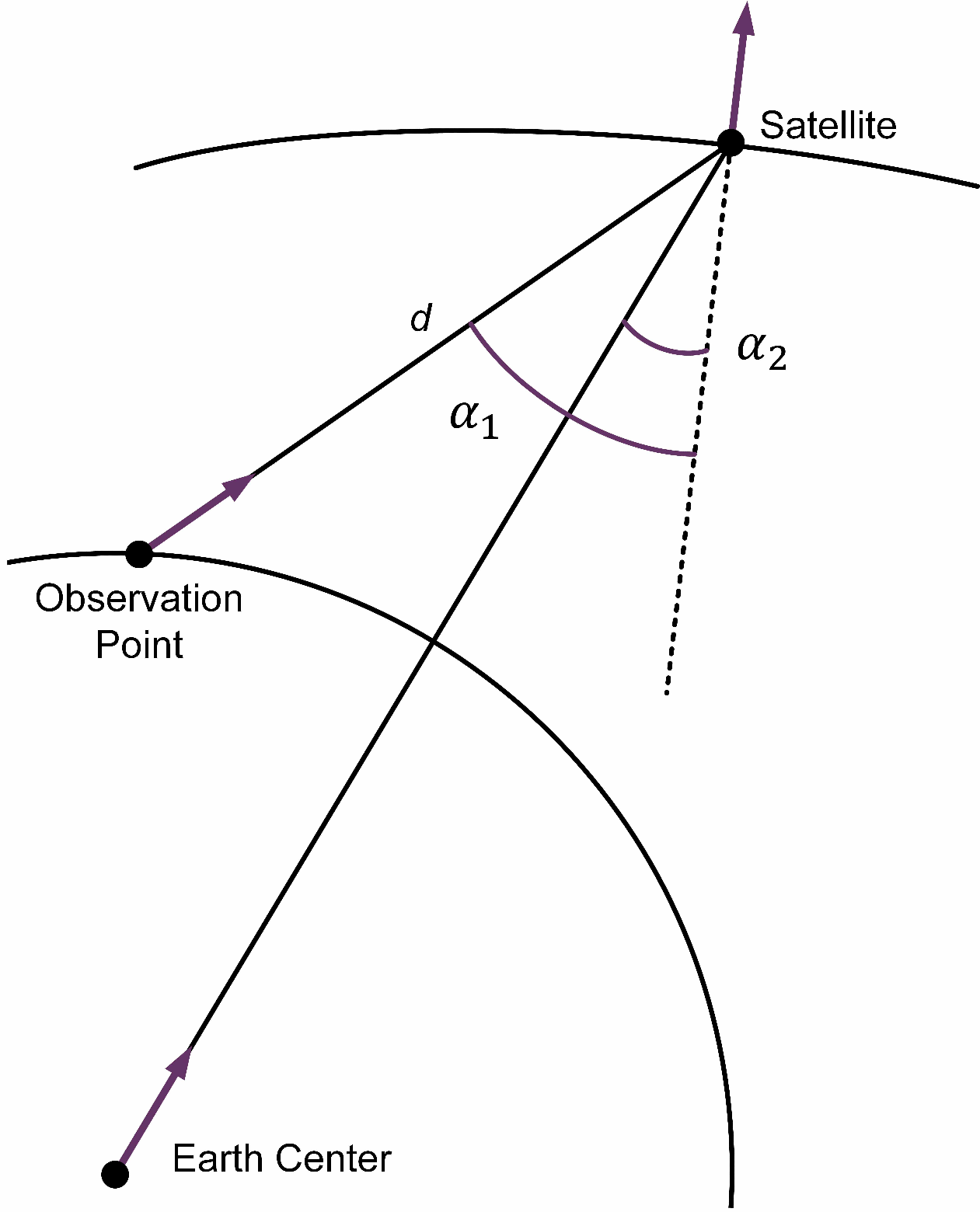}
      \caption{The $\alpha_2$ represents the angle between the satellite-to-Earth center vector and the normal vector to the satellite orbital surface. The term $\alpha_1$ represents the angle between the satellite-to-observation point vector and the normal vector to the satellite orbital surface, $d$ is the distance between the satellite and the observation point.}
         \label{aa_fig8}
 \end{figure}
\end{appendix}

\end{document}